\documentclass[useAMS,usenatbib]{mn2e}

\usepackage{times}

\newcommand{\beq}{\begin{equation}}
\newcommand{\eeq}{\end{equation}}
\newcommand{\lab}{\label}

\newcommand{\bfxi}{\mbox{\boldmath $\xi$}}

\newcommand{\bfomega}{\mbox{\boldmath $\omega$}}
\newcommand{\bftheta}{\mbox{\boldmath $\theta$}}

\newcommand{\bfalpha}{\mbox{\boldmath $\alpha$}}
\newcommand{\bfbeta}{\mbox{\boldmath $\beta$}}
\newcommand{\bfgamma}{\mbox{\boldmath $\gamma$}}

\newcommand{\bfx}{\mathbf{x}}

\newcommand{\bfv}{\mathbf{v}}

\begin{document}

\title[On the motion of a gravitational lens]{On gravitational lensing by deflectors in motion}
\author[M. Sereno]{M. Sereno$^{1,2,3}$\thanks{E-mail:
mauro.sereno@na.infn.it}
\\
$^{1}$Dipartimento di Scienze Fisiche, Universit\`{a} degli Studi di
Napoli `Federico II', Via Cinthia, Monte S. Angelo, 80126 Napoli,
Italia
\\
$^{2}$Istituto Nazionale di Astrofisica - Osservatorio Astronomico di
Capodimonte, Salita Moiariello, 16, 80131 Napoli, Italia
\\
$^{3}$Istituto Nazionale di Fisica Nucleare, Sez. Napoli, Via Cinthia,
Monte S. Angelo, 80126 Napoli, Italia}

\date{January 10, 2005}

\maketitle

\begin{abstract}
Gravitational lensing by a spinning deflector in translational motion
relative to the observer is discussed in the weak field, slow motion
approximation. The effect of rotation, which generates an intrinsic
gravito-magnetic field, separates from that due to radial motion.
Corrections to the lens equation, deflection angle and time delay are
derived.
\end{abstract}


\section{Introduction}

Gravitational lensing is a well recognized practical tool in modern
astrophysics. It is is one of the most deeply investigated phenomena
of gravitation and progressive development of technological
capabilities demands to lead a full analysis on the basis of
higher-order effects \citep{ser03prd}. Corrections due to the motion
of the deflector deserve particular attention on both theoretical and
phenomenological sides. The motion of the lens might play a role in
various astrophysical systems
\citep{fri03a,ser03hom,se+ca02,ser04tmd}. The measurement of an effect
on time delay of luminous signals, due to a translational motion of
the deflector, was recently claimed in \citet{fo+ko03} but a debate
about correct theory and analysis behind the experiment is still under
way \citep{sam04,kop05}. Besides translational motion on a static
background, mass-energy currents relative to other masses generate
space-time curvature. This phenomenon, known as intrinsic
gravito-magnetism, is a new feature of general theory of relativity
and other conceivable theories of gravity and is still waiting for an
experimental confirmation \citep{ci+wh95}.

Various authors have investigated the problem of light deflection by
lenses in motion. The effect of a translational motion of the
deflector on a static background has been widely discussed
\citep{py+bi93,ko+sc99,fri03a,fri03b,hey04,wu+sp04} and an early
disagreement on the correction factor of first order in velocity has
been solved \citep[see][and references therein]{wu+sp04}.
Gravitational lensing by spinning deflectors has been also addressed
with very different approaches
\citep{ep+sh80,ib+ma82,iba83,dym86,gli99,ciu+al03,ser04tmd,ser02pla,ser03prd,ser03hom,se+ca02}.
The two phenomena are really different since the intrinsic angular
momentum of a body cannot be generated or eliminated by a Lorentz
transformation. Whereas the effect due to a translational motion is a
consequence of the existence of the standard gravito-electric field
plus local invariance, the problem of light deflection by a deflector
with angular momentum is related to the existence of the dragging of
inertial frames and of the related gravito-magnetic field. In order to
show that intrinsic gravito-magnetism and the local Lorentz invariance
on a static background are two fundamentally different phenomena,
space-time curvature by mass-energy currents relative to other mass
can be precisely characterized by a frame- and coordinate-independent
method, based on space-time curvature invariants \citep[see][Section
6.11]{ci+wh95}.

In this letter, we address the problem of light deflection by an
extended lens with angular momentum in translational motion on a
static background. To this aim, we adopt the usual framework of
gravitational lensing theory, i.e. {\it i)} weak field and slow motion
approximation for the lens and {\it ii)} thin lens hypothesis
\citep{sef,pet+al01}. The paper is as follows. In Sec.~\ref{stat},
gravitational lensing in a stationary space-time is reviewed. In
Sec.~\ref{time}, we discuss the combined effect of a
roto-translational motion of the deflector on time delay.
Section~\ref{lens} is devoted to the discussion of the bending angle
and to the writing of the lens equation. Section~\ref{conc} contains
some final considerations.

\section{Stationary spinning lenses}
\label{stat}

We are interested in the gravitational field by a weak deflector in
slow motion. Matter velocities are much less than $c$, the speed of
light in the vacuum, and matter stresses are also small. The metric is
asymptotically flat and deviates only slightly from the Minkowski one.
Up to leading order in $c^{-3}$, such a metric can be written in
Cartesian coordinates $x^\alpha
\equiv \left\{ t, \bfx\right\}$\footnote{Greek indeces run from 0 to 3
whereas Latin ones from 1 to 3.} as
\beq
\lab{wf1}
ds^2 \simeq \left( 1+2\frac{\phi}{c^2}\right)c^2dt^2-8c dt \frac{
\mathbf{V} {\cdot} d \bfx}{c^3}- \left( 1-2\frac{\phi}{c^2}\right)d\bfx {\cdot} d
\bfx ,
\eeq
where a dot represents the Euclidean scalar product between two
three-dimensional vectors. Let us consider a stationary space-time.
Here, $\phi$ reduces to the Newtonian potential,
\beq
\lab{wf2}
\phi_\mathrm{s} (\bfx) \simeq - G \int_{\Re^3} \frac{\rho (\bfx^{'} ) }{ | \bfx -
\bfx^{'} |} d^3 x^{'},
\eeq
where $\rho$ is the mass density and $G$ is the gravitational
constant, and $\mathbf{V}$ is a vector potential taking into account
the gravito-magnetic field produced by mass currents,
\beq
\lab{wf3}
\mathbf{V}_\mathrm{s} (\bfx) \simeq -G \int_{\Re^3}
\frac{ ( \rho \bfv ) (\bfx^{'})  }{ | \bfx -\bfx^{'}| } d^3 x^{'}
\eeq
where $\bfv$ is the velocity field of the mass elements of the
deflector. In terms of an angular velocity $\bfomega$, $\bfv =
\bfomega {\times} \bfx$.

In a conformally stationary space-time, actual light rays can be
determined through the Fermat's principle \citep{sef}, which states
that light traverses a path whose optical length is stationary
compared with neighbouring paths \citep{ros57}. A curved space-time
embedded with a stationary metric can be interpreted as a flat one
with an effective index of refraction, $n_\mathrm{s}$
\citep{sef,ser03prd}. A light signal emitted at the source will arrive
after
\beq
\lab{wf5}
\Delta T =\frac{1}{c}\int_\gamma n_\mathrm{s}  d l,
\eeq
where $\gamma$ is the spatial projection of the light curve. The
Fermat's principle can be expressed as
\beq
\delta \int n_\mathrm{s} d l =0 .
\eeq

For a stationary, slowly rotating, weak lens, the effective refraction
index, $n_\mathrm{s}$, is given by \citep{sef,ser03prd}
\beq
\lab{wf4}
n_\mathrm{s} \simeq 1-\frac{2}{c^2} \phi_\mathrm{s} +\frac{4}{c^3}
\mathbf{V}_\mathrm{s} {\cdot} \mathbf{e},
\eeq
where $\mathbf{e} \equiv d \bfx/d l$ is the unit tangent vector of a
ray and $d l \equiv \sqrt{ \delta_{ij}d x^i d x^j}$ is the Euclidean
arc length. The total travel time is \citep{ser02pla,ser03prd}
\beq
\label{stat1}
\Delta T_\mathrm{s} \simeq \Delta T_\mathrm{s}^\mathrm{geo}+
\Delta T_\mathrm{s}^\mathrm{Sh}
+ \Delta T_\mathrm{s}^\mathrm{GRM}
\eeq
where $T_\mathrm{s}^\mathrm{geo}$ is the geometrical time delay,
$\Delta T_\mathrm{s}^\mathrm{Sh}$ is the Shapiro time delay, i.e. the
gravitational delay at the post-Newtonian order,
\beq
\Delta T_\mathrm{s}^\mathrm{Sh} \equiv  - \frac{2}{c^3} \int_\gamma \phi_\mathrm{s} dl
\eeq
and $\Delta T_\mathrm{s}^\mathrm{GRM}$ is the gravito-magnetic time
delay
\beq
\Delta T_\mathrm{s}^\mathrm{GRM} \equiv \frac{4}{c^4} \int_\gamma
\mathbf{V}_\mathrm{s}{\cdot} \mathbf{e} dl .
\eeq
As usual, the lens is assumed to be thin and weak. The actual path of
the photon deviates negligibly from the undeflected path. It is useful
to employ the spatial orthogonal coordinates $(\xi_1, \xi_2, l )$,
centred on the lens and such that the $l$-axis is along the incoming
light ray direction $\mathbf{e}_{\rm in}$ and the vector $\bfxi$ spans
the lens plane. So, in the integrand functions, we can use $\bfx
= l \mathbf{e}_\mathrm{in}+ \bfxi_\mathrm{in} $.
The unperturbed photon impacts the lens plane in $\bfxi_\mathrm{in}$.
The main contribution to the potential time delay is \citep{sef}
\beq
\lab{pot2}
\Delta T_\mathrm{s}^\mathrm{Sh} \simeq
-\frac{4 G}{c^3}\int_{\Re^2}d^2\xi^{'}\Sigma(\bfxi^{'})
\ln \frac{|\bfxi -\bfxi^{'}|}{\xi_0},
\eeq
where $\xi_0$ is a length scale in the lens plane and $\Sigma$ is the
projected surface mass density of the deflector,
\beq
\lab{pot3}
\Sigma(\bfxi)\equiv \int_\mathrm{los} \rho(\bfxi,l)\ dl;
\eeq
the gravito-magnetic correction to the potential time delay, up to the
order $v/c$, can be expressed as \citep{ser02pla}
\beq
\Delta T_\mathrm{s}^\mathrm{GRM} \simeq
\frac{8 G}{c^4} \int_{\Re^2} d^2 \xi^{'} \Sigma(\bfxi^{'})
\langle \mathbf{v}{\cdot}\mathbf{e}_\mathrm{in} \rangle_\mathrm{los} (\bfxi^{'})\ln
\frac{|\bfxi -\bfxi^{'}|}{\xi_0},
\eeq
where $\langle \mathbf{v}{\cdot}\mathbf{e}_\mathrm{in}\rangle_\mathrm{los}$
is the weighted average, along the line of sight $\mathbf{
e}_\mathrm{in}$, of the component of the velocity $\bf v$ along
$\mathbf{e}_\mathrm{in}$,
\beq
\lab{pot4}
\langle \mathbf{v}{\cdot}\mathbf{e}_\mathrm{in}\rangle_\mathrm{los} (\bfxi)
\equiv \frac{\int (\mathbf{v}(\bfxi,l){\cdot} \mathbf{e}_\mathrm{in})
\ \rho(\bfxi,l)\ dl}{\Sigma(\bfxi)}.
\eeq
In the thin lens approximation, the only components of the velocities
parallel to the line of sight enter the equations of gravitational
lensing. We remind that the time delay function is not an observable,
but the time delay between two actual rays can be measured. In the
above expressions for the time delays, we have neglected some not
relevant additive constants.

\section{Time delay by shifting and spinning lenses}
\label{time}

We want now to consider an additional translational motion of the
deflector. The corresponding metric can be obtained by applying a
coordinate transformation to the metric of the same deflector with the
center of mass at rest, which is expressed by Eq.~(\ref{wf1}) with the
potentials in Eqs.~(\ref{wf2},~\ref{wf3}). We limit to a motion with
slow velocity, $u \sim {\cal O}(v)$, and negligible acceleration.
Hereafter, in this section, the primed coordinates will refer to the
rest frame. A rigid motion along a path $\bfgamma$ can be accounted
for by a change of coordinates $\bfx^{'} \rightarrow
\bfx = \bfx^{'}+\bfgamma (t^{'})$ \citep{fri03b}. Limiting to negligible
acceleration of the path, the change in the time coordinate is such
that $ dt= dt^{'} +\dot{\bfgamma}{\cdot} d\bfx^{'}/c^2$ \citep{fri03b}. The
metric takes the same form in Eq.~(\ref{wf1}) with
\begin{eqnarray}
\phi (\bfx, t) & = & \phi_\mathrm{s} (\bfx -\bfgamma(t) ) \label{movi1} \\
\mathbf{V} (\bfx, t) & = & \mathbf{V}_\mathrm{s} (\bfx -\bfgamma(t) ) +
\frac{\mathbf{u} }{c} \phi_\mathrm{s} (\bfx -\bfgamma(t) ), \label{movi2}
\end{eqnarray}
where $\mathbf{u} \equiv \dot{\bfgamma}$.

The metric in Eq.~(\ref{wf1}) with the potentials given in
Eq.~(\ref{movi1},~\ref{movi2}) is no more stationary and Fermat's
principle cannot be applied as done in Sec.~(\ref{stat}). However,
under suitable assumptions, the computation of the gravitational
lensing quantities can proceed nearly in the same way. In fact, in
usual astrophysical systems, the transit time through the deflector
can be considered small with respect to the total travel time
\citep{sef}. We can assume that interaction happens instantaneously at
the moment when the photon passes the lens, $t_\mathrm{d}$. The
position of the center of mass of the lens at $t_\mathrm{d}=0$ locates
the origin of the coordinate system. During the transit time, the
velocities are approximately constant. So, we can consider the
components of the metric tensor fixed at $t=t_\mathrm{d}$ and proceed
as in the stationary case. The corresponding ad hoc refraction index
is written solving for $dt$ the equation $ds^2=0$, and properly
considering the difference between the proper arc-length in a curved
space-time and the Euclidean arc length \citep{ser03prd}. We obtain
\begin{eqnarray}
n & \simeq & 1-\left( 1-2 \frac{ u_\mathrm{los}(t_\mathrm{d}) }{c}
\right)
\frac{ 2 \phi_\mathrm{s}(\bfx -\bfgamma(t_\mathrm{d}) )}{c^2} \nonumber \\
& + &
\frac{4}{c^3} V_\mathrm{s,los} (\bfx -\bfgamma(t_\mathrm{d}) ) \label{movi3},
\end{eqnarray}
where again the subscript los denotes the component of a vector along
the line of sight. The index of refraction in Eq.~(\ref{movi3}) can be
also obtained from the index in the stationary case, through a Lorentz
transformation. As first demonstrated by Fizeau in the classic
experiment of light traversing a moving fluid, it is
\beq
\label{movi4}
\frac{1}{n} \simeq \frac{1}{n_\mathrm{s}} - \frac{u \cos \vartheta}{c}
\left( 1 - \frac{1}{n_\mathrm{s}^2 } \right),
\eeq
where $\vartheta$ is the angle between the direction of propagation of
the light and the velocity of the medium, i.e. $u \cos \vartheta =
u_\mathrm{los}$.

The computation of the time-delay can be performed by defining a new
integration variable, $q$ \citep{fri03a},
\beq
q \equiv l - \mathbf{e}_\mathrm{in} {\cdot} \bfgamma (t_\mathrm{d} );
\eeq
in terms of this variable
\beq
\label{movi7}
\bfx - \gamma(t_\mathrm{d}) = q \mathbf{e}_\mathrm{in}+ \bfxi_\mathrm{in} -
\mathbf{u}_\perp t_\mathrm{d}
\eeq
where $\mathbf{u}_\perp$ is the projected velocity of the deflector in
the lens plane, i.e. transverse to the line of sight, and
\beq
\label{movi8}
dq = \left( 1- u_\mathrm{los}/c \right) dl .
\eeq
The time delay by a moving, spinning deflector can be expressed in
terms of the same stationary deflector. By changing the integration
variable, we find
\beq
\label{movi9}
\Delta T \simeq \Delta T^\mathrm{geo} - \left( 1- \frac{u_\mathrm{los} }{c}
\right ) \int_\gamma \frac{2 \phi_\mathrm{s}}{c^3} dq
+\int_\gamma \frac{4\mathbf{V}_\mathrm{s}}{c^4} dq .
\eeq
Solving the integrals under the hypothesis of a thin lens, we get
\begin{eqnarray}
\Delta T \simeq \Delta T_\mathrm{s}^\mathrm{geo}
& + & \left( 1- \frac{u_\mathrm{los} }{c} \right )\Delta
T_\mathrm{s}^\mathrm{Sh} ( \bfxi_\mathrm{in} - \mathbf{u}_\perp
t_\mathrm{d})
\nonumber
\\
& +& \Delta T_\mathrm{s}^\mathrm{GRM} ( \bfxi_\mathrm{in} - \mathbf{
u}_\perp t_\mathrm{d})
\end{eqnarray}
The translational motion of the lens enters in two distinct ways. The
radial motion affects the Shapiro time delay through an overall
multiplicative factor, so that the correction to the time delay due to
the transversal motion is $\Delta T^\mathrm{tra} =
-( u_\mathrm{los}/c ) \Delta T_\mathrm{s}^\mathrm{Sh}$. The transverse
motion affects only the relative angular separation between the lens
and the source, i.e the impact parameter, which is now time dependent.

The weak field hypothesis and slow motion approximation are enough to
face almost all astrophysical systems \citep{sef}. In fact, peculiar
velocities of high redshift galaxies are only a tiny perturbations on
the Hubble flow, of order of $\sim 10^{-3}c$. Let us consider a
background quasar which is lensed by a foreground galaxy. We can model
the lens as a singular isothermal sphere \citep{ser04tmd}. An
approximate relation holds between the Shapiro time delay and the
gravito-magnetic time delay \citep{ser04tmd},
\beq
\Delta T^\mathrm{GRM} \simeq -9 \sin \gamma \left( \frac{\sigma_\mathrm{v}}{c}
\right) \lambda \Delta T_\mathrm{s}^\mathrm{Sh},
\eeq
where $\gamma$ is the angle between the projection of the rotation
axis in the plane of the sky and the line trough the two nearly
collinear images, $\lambda$ is the spin parameter, giving an estimate
of the total angular momentum of the lens, and $\sigma_\mathrm{v}$ is
the velocity dispersion of the galaxy. Corrections due to rotation and
translational motion are nearly of the same order for usual systems.
For a typical lensing galaxy at $z_{\rm d}=0.5$ with
$\sigma_\mathrm{v} \sim u_\mathrm{los}
\sim 10^{-3}$, and a background source at $z_{\rm s} = 2.0$, the time
delay is $\sim 200$~days for a source nearly in the middle of the
Einstein radius, whereas the corrections are $\Delta T^\mathrm{tra}
\sim 0.2$~days and $\Delta T^\mathrm{GRM}
\sim 0.1$-$ 0.2 {\times} \sin \gamma$~days for $\lambda \sim 0.05$-0.1.

\section{Lens equation}
\label{lens}

Just as for a stationary space-time, where the Fermat's principle
exactly holds, even if the lens is in slow motion, the bending angle
$\bfalpha$ and the gravitational time delay, $\Delta
T^\mathrm{pot}=\Delta T -\Delta T^\mathrm{geo}$, can be related by a
gradient \citep{fri03b},
\beq
\label{lens1}
\bfalpha = -\nabla_\perp (c \Delta T^\mathrm{pot}),
\eeq
where $\nabla_\perp \equiv \nabla-\mathbf{e}_\mathrm{in} (\mathbf{
e}_\mathrm{in} {\cdot} \nabla) $. The bending angle turns out to be
\begin{eqnarray}
\bfalpha (\bfxi ) & \simeq  & \frac{4 G}{c^2}\int_{\Re^2}d^2\xi^{'} \left\{
\frac{\bfxi -\bfxi^{'}}{|\bfxi -\bfxi^{'}|^2}
\Sigma(\bfxi^{'} -\mathbf{u}_\perp t_\mathrm{d} ) \right.  \label{lens2} \\
 &  {\times} & \left. \left( 1-
\frac{ u_\mathrm{los}}{c}+\frac{2}{c}
\langle \mathbf{v}{\cdot}\mathbf{e}_\mathrm{in} \rangle_\mathrm{los}(\bfxi^{'}
-\mathbf{u}_\perp t_\mathrm{d})\right) \right\} \nonumber.
\end{eqnarray}
Once again, three terms contribute to $\bfalpha$. The main term is due
to the static, gravito-electric field of the corresponding lens at
rest. The second term derives from the spin and take the same form as
for a stationary lens with angular momentum. The third effect on the
bending angle, due to a translational motion of the deflector,  can be
interpreted in terms of standard aberration of light in an optically
active medium with an effective index of refraction induced by the
gravitational field of the lens \citep{fri03b}. The bending angle by a
point-like moving lens carries a pre-factor $(1 - u_\mathrm{los}/c)$
with respect to the bending angle in the static case \citep{fri03b}.
For a stationary spinning deflector, the path of the light-ray does
not stay on a plane \citep{ser03hom}, as it is for a spherically
symmetric lens. However, since in our approximations, deviations
induced by the angular momentum of the lens are small, the same
arguments based on aberration can be applied as in \citet{fri03b},
providing an independent confirmation to Eq.~(\ref{lens2}) as far as a
radial motion is concerned.

Let us finally write the lens equation. We introduce the angular
position of the source in the plane of the sky, $\bfbeta$, and the
angular position of the image in the lens plane $\bftheta =
\bfxi /D_\mathrm{d}$, where $D_\mathrm{d}$ is the angular diameter
distance from the observer to the deflector. For a system of
point-like lenses, velocity effects do not alter the form of the lens
equation with respect to the static case \citep{ko+sc99,hey04}. For
the general case of a spinning and shifting deflector, the lens
equation takes the form,
\beq
\label{lens3}
\bfbeta = \bftheta - \frac{D_\mathrm{ds}}{D_\mathrm{s}}\bfalpha (\bftheta),
\eeq
where and $D_\mathrm{s}$ and $D_\mathrm{ds}$ are the angular diameter
distances from the observer to the source and from the deflector to
the source, respectively. With respect to the static case, the
position of the lens is time dependent and the bending angle is
corrected for factors due to the radial motion and the angular
momentum of the deflector.

In the limit of slow velocities, translational and rotational motions
are separated and add together to correct lensing quantities. In
general, this is not true. To first order in deflection, the
deflection angle for a lens with a relativistic translational motion
can be directly obtained from the stationary deflection angle caused
by a lens at rest by applying a Lorentz transformation
\citep{wu+sp04}. The stationary deflection angle can be written as a
sum of the gravito-electric and the gravito-magnetic terms,
$\bfalpha^\mathrm{GRE}+\bfalpha^\mathrm{GRM}$. For a relativistic
translational velocity,
\beq
\label{lens4}
\bfalpha (\mathbf{u})= \sqrt{ \frac{1-u_\mathrm{los}/c}{1+u_\mathrm{los}/c} }
\left( \bfalpha^\mathrm{GRE}+\bfalpha^\mathrm{GRM} \right)
\eeq
The scaling factor in Eq.~(\ref{lens4}), which reduces to
$1-u_\mathrm{los}/c$ for a slow motion, applies in the same way to the
main term and to the gravito-magnetic correction.

\section{Concluding remarks}
\label{conc}

We have considered the effects on gravitational lensing of both the
translational motion on a static background and of an intrinsic
angular momentum of the deflector. In the weak-field, slow motion
approximation the effects on bending and time delay of light rays are
separated and adds together. Only the radial motion of the centre of
mass enters the corrections, whereas the transverse motion has to be
considered only when determining the lens position and the relative
separation between lens and source. The radial and the angular
velocity enter the corrections with a different scaling. The internal
rotational velocity must be weighted along the line of sight and
carries an extra factor 2 with respect to the radial velocity, that
must be evaluated at the time the photon passes the lens. This shows
that results in \citet{ser02pla,se+ca02}, which refer to spinning
lenses, are correct, differently from what claimed in \citet{wu+sp04}.
However, results in \citet{ser02pla} can not be applied to a radial
motion.

Thanks to usual approximations, our analysis does not limit to
point-like lenses and extended deflector have been considered. The
internal velocity field enters the equations. This allows to solve the
question whether the gravito-magnetic correction to the deflection of
light can probe the inner structure of a lens \citep{wu+sp04}. Even if
the main contribution for a light ray propagating outside the lens is
due to the total angular momentum, light rays crossing the inner
structure of the deflector probe only a ``partial" angular momentum.

\section*{Acknowledgments}
I wish to thank an anonymous referee for the detailed suggestions.

\bibliographystyle{mn2e}
\bibliography{ME1225Lrv}

\end{document}